\documentclass[12pt]{article}
\pdfoutput=1
\usepackage{bbm,latexsym,amsmath,amssymb}		
\usepackage[colorlinks=true,linkcolor=redLinks,citecolor=greenLinks,urlcolor=redLinks, pdfborder={0 0 1}]{hyperref} 
\usepackage{booktabs}
\usepackage{graphicx}
\usepackage{color}
\usepackage{amsmath}
\usepackage{amssymb}
\usepackage{mathrsfs}
\usepackage{bbold}

\definecolor{greenLinks}{rgb}{0, 0.6, 0} 
\definecolor{blueLinks}{rgb}{0, 0, 0.6}
\definecolor{redLinks}{rgb}{0.6, 0, 0} 
\renewcommand{\thefootnote}{\fnsymbol{footnote}}
\allowdisplaybreaks
\begin{document}
\title{
\normalsize \hfill UWThPh-2014-22 \\
\normalsize \hfill IFIC/14-68 \\*[8mm]
\LARGE Classification of lepton mixing patterns\\
 from finite flavour symmetries}

\author{
Renato M. Fonseca\,$^{(1)}$\thanks{E-mail: renato.fonseca@ific.uv.es; contribution to the proceedings of the \textit{37th International
Conference on High Energy Physics} (ICHEP 2014), Valencia, Spain,
2 -- 9 July 2014, based on \cite{Fonseca:2014koa}}
\setcounter{footnote}{2}
and\,
Walter Grimus\,$^{(2)}$\thanks{E-mail: walter.grimus@univie.ac.at} \
\\[5mm]
$^{(1)} \! $
\small AHEP Group, Instituto de F\'isica Corpuscular,
  C.S.I.C./Universitat de Val\`encia \\ 
\small Edificio de Institutos de Paterna, Apartado 22085, E--46071
  Val\`encia, Spain \\ 
\\[2mm]
$^{(2)} \! $
\small University of Vienna, Faculty of Physics \\
\small Boltzmanngasse 5, A--1090 Vienna, Austria
\\[2mm]
}
\date{October 15, 2014}

\maketitle

\begin{abstract}
Flavour symmetries have been used to constrain both quark and lepton
mixing parameters. In particular, they can be used to completely fix
the mixing angles. For the lepton sector, assuming that neutrinos
are Majorana particles, we have derived the complete list of mixing
patterns achievable in this way, as well as the symmetry groups associated
to each case. Partial computer scans done in the past have hinted
that such list is limited, and this does indeed turn out to be the
case. In addition, most mixing patterns are already 3-sigma excluded
by neutrino oscillation data. 
\end{abstract}
\newpage
\renewcommand{\thefootnote}{\arabic{footnote}}
\section{Introduction}




\section{\label{section1}Residual symmetries of the lepton mass matrices}

Two lepton mixing angles are known to be large, yet there is no established
theory which explains their values. One hypothesis is that the lepton
flavour structure is explained by the existence of some finite symmetry
which acts on flavour space. Given that symmetry is known to play
a critical role in fundamental physics, this is a particularly appealing
possibility, which was explored extensively in the literature. However,
most research has focused on building models which incorporate
particular flavour symmetries, with little attention being given
to the problem of systematically cataloging the various possibilities.

References \cite{Holthausen:2012wt,Parattu:2010cy} are notable exceptions, where  the GAP program \cite{GAP} was used to make scans over various groups and representations, yielding lists of lepton mixing matrices which can be obtained from finite flavour symmetries. However, even though these are interesting pioneering works in this area, by the very nature of such computational scans, they are necessarily partial since there are infinitely many groups to be searched. Furthermore, it is useful to have not just a computational understanding of the problem, but an analytical one as well.

In \cite{Fonseca:2014koa} we have done such a systematic study for the cases where
the lepton mixing matrix is predicted completely by a finite discrete
symmetry, leaving no relation between mixing angles and mass ratios.
It is then appropriate to first briefly review the relation between
this class of models and the residual symmetries of the lepton mass
matrices. Assuming that neutrinos are Majorana particles,
the mass terms of leptons are the following:
\begin{align}
\mathscr{L}_{\textrm{mass}} & =-\overline{\ell}_{L}M_{\ell}\ell_{R}+\frac{1}{2}\nu_{L}^{T}C^{-1}M_{\nu}\nu_{L}+\textrm{h.c.}
\end{align}
A transformation $\left(\ell_{L},\nu{}_{L}\right)\rightarrow\left(T\ell_{L},S_{i}\nu_{L}\right)$
with $T=U^{\dagger} \cdot \textrm{diag}\left(\lambda_{1}^{(0)},\lambda_{2}^{(0)},\lambda_{3}^{(0)}\right) \cdot U$
and $S_{1,2,3}$ given by $\textrm{diag}\left(-1,+1,+1\right)$, $\textrm{diag}\left(1,-1,+1\right)$,
$\textrm{diag}\left(+1,+1,-1\right)$ will impose $U$ as the lepton
mixing matrix, provided that the eigenvalues of $T$ are distinct.%
\footnote{If this is not the case, two $T$ matrices are needed.%
} To be precise, given that this symmetry leaves the lepton masses
completely unconstrained, one is able to impose that the absolute
values of the entries of the mixing matrix are given by $\left|U\right|_{ij}\equiv\left|U_{ij}\right|$,
up to row and column permutations of $U$.

The $\left\{T,S_{i}\right\} $ matrices generate a group $G$, which
is infinite in general. However, it should be noted that the cases where
$G$ is finite are particularly interesting and deserve special attention, 
as they provide a way to avoid having massless scalar particles. One can ask:
 when does $T$ and the $S_{i}$
generate a finite group?

The answer to this question depends on both $U$ and the eigenvalues
of the $T$ matrix (the $\lambda_{i}^{(0)}$). For example, in \cite{Lam:2008sh} it was proven that for tribimaximal
mixing (TBM) --- which was an acceptable ansatz until the reactor angle was measured to be appreciable --- the resulting group is finite if and only if
the eigenvalues of $T$ are $\left(1,\omega,\omega^{2}\right)$ up to permutations of these values and multiplication by a common root
of unity ($\omega\equiv \exp(2\pi i/3)$). Depending on this multiplicative root of unity, $G$ might
be $S_{4}$ or some other group which contains it as a subgroup. It
is therefore natural to associate tribimaximal mixing with the effective
symmetry group $S_{4}$.

The technique used in \cite{Lam:2008sh} to derive
this connection relied on scanning all finite subgroups of $SU(3)$
(as listed in \cite{Fairbairn:1964sga,Bovier:1980gc}; see also \cite{Grimus:2013apa}) and on analyzing their 3-dimensional irreducible
representations. However, there is a simpler way to derive the same result.

\section{Roots of unity, and finite groups}

The finiteness of a group $G$ requires that the eigenvalues of its
elements $g\in G$, in any representation, must be roots of unity.
This is a direct consequence of the fact that there is always some
positive integer $n$ such that $g^{n}=e$ (the identity element).
As such, the eigenvalues of $T$ (the $\lambda_{i}^{(0)}$) as
well as those of $TS_{j}$ (which we may call $\lambda_{i}^{(j)}$)
must all be roots of unity. Computing the eigenvalues of the $TS_{j}$
matrices as a function of those of $T$ can be done, but the expressions
are in general complicated so, instead of looking at the $\lambda_{i}^{(j)}$
individually, we may consider the combinations $\lambda_{1}^{(j)}+\lambda_{2}^{(j)}+\lambda_{3}^{(j)}$
which are just the traces of the matrices $TS_{j}$ (no additional
useful information is obtainable from $\lambda_{1}^{(j)}\lambda_{2}^{(j)}+\lambda_{1}^{(j)}\lambda_{3}^{(j)}+\lambda_{2}^{(j)}\lambda_{3}^{(j)}$
and $\lambda_{1}^{(j)}\lambda_{2}^{(j)}\lambda_{3}^{(j)}$). For $U=U_{TBM}$, it is easy to check that
\begin{align}
\lambda_{1}^{(0)}+\lambda_{2}^{(0)}+\lambda_{3}^{(0)}+3\left(\lambda_{1}^{(2)}+\lambda_{2}^{(2)}+\lambda_{3}^{(2)}\right) & =0\,.
\end{align}
This is a vanishing sum of roots of unity with integer coefficients,
whose solutions can be found using theorem 6 of \cite{Conway_Jones:1976}.
There are two: either (a) $\lambda_{1}^{(0)}=\lambda_{2}^{(0)}=\lambda_{3}^{(0)}=-\lambda_{i}^{(2)}$
and $\lambda_{j}^{(2)}=-\lambda_{k}^{(2)}$ ($i\neq j\neq k\neq i$), or (b) $\left(\lambda_{1}^{(0)},\lambda_{2}^{(0)},\lambda_{3}^{(0)}\right)=\xi\left(1,\omega,\omega^{2}\right)$
and $\left(\lambda_{1}^{(2)},\lambda_{2}^{(2)},\lambda_{3}^{(2)}\right)=\xi'\left(1,\omega,\omega^{2}\right)$
for some roots of unity $\xi$, $\xi'$ (up to permutations amongst
the $\lambda_{i}^{(0)}$/$\lambda_{i}^{(2)}$). Solution (a) implies
that the eigenvalues of $T$ are completely degenerate ($T\propto\mathbb{1}$)
which is unacceptable from the physical point of view (see section \ref{section1}).
Solution (b) is the one reported in \cite{Lam:2008sh}.

\section{The lepton mixing matrices compatible with finite symmetries}

Instead of just scanning the allowed $\lambda_{i}^{(0)}$ for a particular
value of $U$, in \cite{Fonseca:2014koa} we considered the problem in full generality,
which resulted in the list of all the values of $\left(\lambda_{i}^{(0)},U\right)$
which lead to a finite group. To do so, we applied known results concerning
roots of unity (see also \cite{Grimus:2013rw}), without having to rely on the classification of the
finite subgroups of $SU(3)$, nor did we have to use group or representation
theory. This is particularly relevant given that finding the finite subgroups of $SU(3)$ and their representations over the last century was not a straightforward task (see \cite{Ludl:2011gn}).

In a first step, it can be shown that, up to row and column permutations,
the representation matrix $X$ of any element of a finite group which
contains the $S_{i}$ is of the form
\begin{alignat}{1}
\left|X\right| & =\left(\begin{array}{ccc}
0 & \frac{1}{\sqrt{2}} & \frac{1}{\sqrt{2}}\\
\frac{1}{\sqrt{2}} & \frac{1}{2} & \frac{1}{2}\\
\frac{1}{\sqrt{2}} & \frac{1}{2} & \frac{1}{2}
\end{array}\right),\left(\begin{array}{ccc}
\frac{1}{2} & \frac{\sqrt{5}-1}{4} & \frac{\sqrt{5}+1}{4}\\
\frac{\sqrt{5}+1}{4} & \frac{1}{2} & \frac{\sqrt{5}-1}{4}\\
\frac{\sqrt{5}-1}{4} & \frac{\sqrt{5}+1}{4} & \frac{1}{2}
\end{array}\right),\nonumber \\ \textrm{ or }
 & \left(\begin{array}{ccc}
1 & 0 & 0\\
0 & \cos\theta & \sin\theta\\
0 & \sin\theta & \cos\theta
\end{array}\right)\,,
\end{alignat}
where $\theta$ is some rational angle ($\theta=\pi q$, with $q\in\mathbb{Q}$).
This applies to $X=T$ but also to $X=T^{2}$, for example. It is
then possible, in a second step, to enumerate all the possible $T$'s
by knowing the allowed values of $\left|T\right|$ and $\left|T^{2}\right|$.

Given that the rows of $U^{*}$ are the eigenvectors of $T$, we
end up with a list of the lepton mixing matrices which are compatible
with a finite flavor group $G$. This list contains 17 sporadic $U$'s
involving three-flavour mixing, as well as the infinite series of
mixing matrices
\begin{align}\label{Eq_infinite_series}
|U|^{2} & =\frac{1}{3}\left(\begin{array}{ccc}
1 & 1+\textrm{Re}\,\rho & 1-\textrm{Re}\,\rho\\
1 & 1+\textrm{Re}\left(\omega\rho\right) & 1-\textrm{Re}\left(\omega\rho\right)\\
1 & 1+\textrm{Re}\left(\omega^{2}\rho\right) & 1-\textrm{Re}\left(\omega^{2}\rho\right)
\end{array}\right)
\end{align}
where $\rho=\exp\left(2\pi ip/n\right)$
can be any root of unity. 
The sporadic cases are associated with the minimal groups $A_{4}$, $S_{4}$,
$A_{5}$, $PSL(2,7)$ and $\Sigma(360\times3)$ and they are excluded by
neutrino oscillation data at more than 3$\sigma$ since they all predict either $\sin^{2}\theta_{13}=0$ or $\sin^{2}\theta_{13}$ equal or larger than $\left(5-\sqrt{21}\right)/12\approx0.035$.

\begin{figure}[tbph]
\begin{centering}
\includegraphics[scale=0.14]{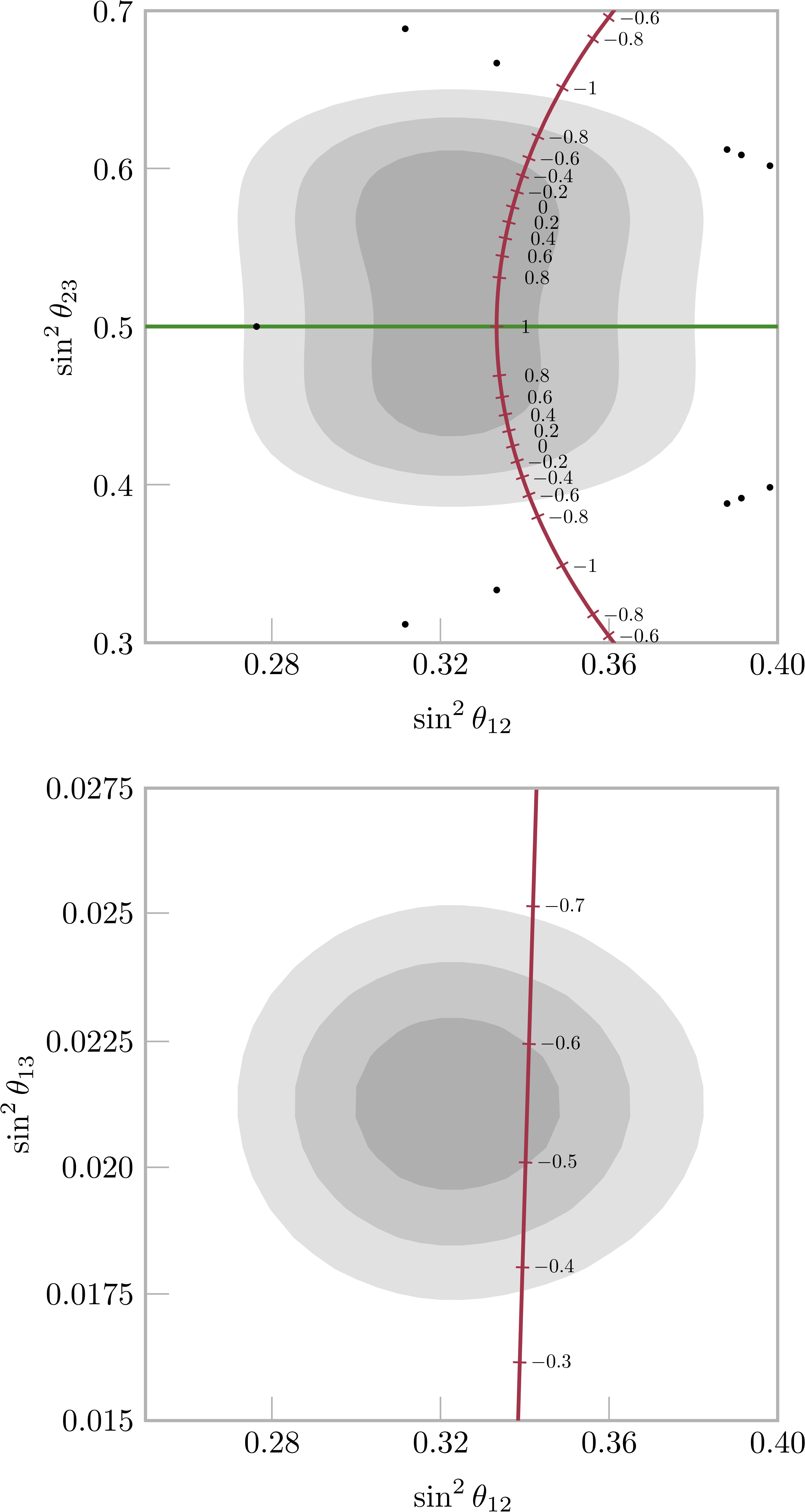}~~~~~~~
\par\end{centering}

\caption{\label{figure1}
Plots showing the angles predicted by some of the mixing matrices
which can be obtained with finite flavour symmetries (figure taken from \cite{Fonseca:2014koa}). The sporadic
mixing matrices appear as dots, while the infinite series of $U$'s
displayed in equation \eqref{Eq_infinite_series} is represented as a red ($\cos^{2}\theta_{13}\sin^{2}\theta_{12}=1/3$)
and a green ($\sin^{2}\theta_{13}=1/3$) line, depending on the column
permutation which is considered. The
numbers along the red curves indicate the values of $\textrm{Re}\left(\rho^{6}\right)$.
The gray areas indicate the 1, 2 and $3\sigma$ regions allowed by
neutrino oscillation experiments, as computed by Forero
et al. \cite{Forero:2014bxa}.
}

\end{figure}

That leaves the infinite series of equation \eqref{Eq_infinite_series}
--- for $-0.69\lesssim\textrm{Re}\left(\rho^{6}\right)\lesssim-0.37$
--- as the only phenomenologically viable case (see figure \ref{figure1}).
Note that this type of mixing implies a $\theta_{23}$ reasonably far from maximal ($\sin^{2}\theta_{23}\sim0.4$ or $0.6$), and a trivial Dirac-type phase $\delta$.
If $n$ is not divisible by 9, it can be shown that the minimal group
$G$ is $\Delta\left(6m^{2}\right)$, with $m=\textrm{lcm}\left(6,n\right)/3$;
otherwise $G$ is a semi-direct product of the form $\left(\mathbb{Z}_{m}\times\mathbb{Z}_{m/3}\right)\rtimes S_{3}$,
with $m=\textrm{lcm}\left(2,n\right)$.

\section*{Acknowledgments}
The work of R.F. was supported by the Spanish grants FPA2011-22975 and Multidark CSD2009-00064 (\textit{MINECO}), PROMETEOII/2014/084 (\textit{Generalitat Valenciana}), and by the Portuguese grants CERN/FP/123580/2011 and EXPL/FIS-NUC/0460/2013 (\textit{Funda\c{c}\~ao para a Ci\^encia e a Tecnologia}).


\begin{thebibliography}{99}

\bibitem{Fonseca:2014koa}
R.~M. Fonseca and W.~Grimus, 
\textit{Classification of lepton mixing matrices from finite
  residual symmetries},
JHEP {\bf 1409} (2014) 033
[arXiv:1405.3678 [hep-ph]].

\bibitem{Holthausen:2012wt}
M.~Holthausen, K.~S.~Lim and M.~Lindner,
\textit{Lepton mixing patterns from a scan of finite discrete groups},
Phys.\ Lett.\ B {\bf 721} (2013) 61
[arXiv:1212.2411 [hep-ph]].

\bibitem{Parattu:2010cy}
K.~M. Parattu and A.~Wingerter,
\textit{Tribimaximal mixing from small groups},
Phys.  Rev. D {\bf 84} (2011) 013011,
[arXiv:1012.2842 [hep-ph]].

\bibitem{GAP}
The GAP~Group,
\textit{GAP -- Groups, Algorithms, and Programming} (2014),
http://www.gap-system.org.

\bibitem{Lam:2008sh}
C.~S.~Lam,
\textit{The unique horizontal symmetry of leptons},
Phys.\ Rev.\ D {\bf 78} (2008) 073015
[arXiv:0809.1185 [hep-ph]].

\bibitem{Fairbairn:1964sga}
W.~Fairbairn, T.~Fulton and W.~Klink,
\textit{Finite and disconnected subgroups of $SU_3$
  and their application to the elementary-particle spectrum}, J. Math. Phys.
  {\bf 5} (1964) 1038.

\bibitem{Bovier:1980gc}
A.~Bovier, M.~{L\"uling} and D.~Wyler,
\textit{Finite subgroups of SU(3)},
 J. Math. Phys. {\bf 22} (1981) 1543.

\bibitem{Grimus:2013apa}
W.~Grimus and P.~O.~Ludl,
\textit{On the characterization of the SU(3)-subgroups of type C and D},
J.\ Phys.\ A {\bf 47} (2014) 075202
[arXiv:1310.3746 [math-ph]].

\bibitem{Conway_Jones:1976}
J.~H. Conway and A.~J. Jones,
\textit{Trigonometric diophantine equations (On vanishing sums of
  roots of unity)},
Acta Arithmetica {\bf 30} (1976) 229.

\bibitem{Grimus:2013rw}
W.~Grimus,
\textit{Discrete symmetries, roots of unity, and lepton mixing},
J.\ Phys.\ G {\bf 40} (2013) 075008
[arXiv:1301.0495 [hep-ph]].

\bibitem{Ludl:2011gn}
P.~O. Ludl,
\textit{Comments on the classification of the finite subgroups of SU(3)},
J. Phys. A: Math. Theor. {\bf 44} (2011) 255204 [Corrigendum: J. Phys. A: Math. Theor. {\bf 45} (2012) 069502]
[arXiv:1101.2308 [math-ph]].

\bibitem{Forero:2014bxa}
D.~V.~Forero, M.~T\'ortola and J.~W.~F.~Valle,
\textit{Neutrino oscillations refitted},
arXiv:1405.7540 [hep-ph].

\end{thebibliography}
\end{document}